%% file: main.tex
\PassOptionsToPackage{final}{graphicx}
\PassOptionsToPackage{final}{listings}
\documentclass[paper]{llncs}
\usepackage[title]{appendix}

\input{packages}

\input{macros}

\title{Online Shielding for Stochastic Systems}
\author{Bettina Könighofer\inst{1,2}  \and 
Julian Rudolf \inst{1}  \and 
Alexander Palmisano\inst{1}  \and  
Martin Tappler\inst{3}  \and  
Roderick Bloem\inst{1}}

\institute{
Graz University of Technology, Institute IAIK, Austria
\and
Silicon Austria Labs, TU-Graz SAL DES Lab, Austria
\and
Schaffhausen Institute of Technology, Schaffhausen, Switzerland
}

\begin{document}

\maketitle

\begin{abstract}

\input{abstract}

\end{abstract}

\setlength{\intextsep}{0pt}%
\section{Introduction}
\input{introduction}

\section{Preliminaries}
\label{sec:preliminaries}
\input{preliminaries}

\section{Setting and Problem Statement}\label{sec:problem}
\input{problem}

\section{Online Shielding for MDPs}
\label{sec:shields}
\input{shields}
\section{Implementation and Experiments}\label{sec:experiments}

\input{experiments}

\section{Conclusion and Future Work}\label{sec:conclusion}
\input{conclusion}

\bibliographystyle{abbrv}
\bibliography{literature}

\end{document}

%% file: packages.tex
\usepackage{hyperref}       
\usepackage{url}            
\usepackage{amsmath,amssymb,amsfonts}       
\usepackage{booktabs}       
\usepackage{xcolor}
\usepackage{xspace}
\usepackage{todonotes}
\usepackage{paralist}
\usepackage[capitalize]{cleveref}
\usepackage{datetime2}
\usepackage{subcaption}
\usepackage{cases}
\usepackage{pgfplots}
\pgfplotsset{compat=newest}
\usepackage{wrapfig}
\usetikzlibrary{fit}

%% file: macros.tex
\newcommand\shieldsym[1][]{%
\raisebox{-1.5pt}{\tikzset{
    shield width/.store in=\shieldwidth,
    shield width=1.5ex,
    shield height/.store in=\shieldheight,
    shield height=1.75ex
}%
\tikz [baseline,#1] \draw (0,\shieldheight) -- (0,\shieldwidth/2) arc [radius=\shieldwidth/2, start angle=-180, end angle=0] -- (\shieldwidth,\shieldheight) -- cycle;%
}}

\newcommand{\Task}{\ensuremath{\textsl{Task}}}

\newcommand{\mcresmin}[2]{\ensuremath{\eta^{\min}_{#1,#2}}}
\newcommand{\mcresmax}[2]{\ensuremath{\eta^{\max}_{#1,#2}}}

\newcommand{\val}[1]{\ensuremath{\textsl{val}_{#1}}}
\newcommand{\optval}[1]{\ensuremath{\textsl{optval}_{#1}}}

\newcommand{\shield}[2]{\ensuremath{\textsl{shield\,}_{#1}^{#2}}}
\newcommand{\shielded}[1]{\ensuremath{#1_{\text{\shieldsym}}}}
\newcommand{\online}[1]{\shielded{#1}}

\newcommand{\snake}{\textsc{Snake}\xspace}

\newcommand{\Distr}{\mathit{Distr}}

\newcommand{\distDom}{X}

\newcommand{\distFunc}{\mu}
\newcommand{\distDomElem}{x}
\newcommand{\supp}{\mathit{supp}}

\newcommand{\R}{\mathbb{R}}

\newcommand{\finally}{\lozenge}

\newcommand{\sinit}{s_{\mathit{I}}} 
\newcommand{\states}[1][]{\mathcal{S}_{#1}}

\newcommand{\dtmc}{\mathcal{D}}
\newcommand{\DtmcInit}[1][]{\ensuremath{\dtmc{#1}=(\mathcal{S}{#1},\sinit{#1},P{#1})}}

\newcommand{\mdp}{\mathcal{M}}

\newcommand{\MdpTupleR}[1][]{\ensuremath{(\mathcal{S}{#1}, s_0, \Act{#1},\pmdp{#1})}}
\newcommand{\MdpInitR}[1][]{\ensuremath{\mdp{#1}=\MdpTupleR[#1]}}

\newcommand{\act}[1][a]{\alpha} 
\newcommand{\Act}{\mathcal{A}}        

\newcommand{\rewFunction}{\ensuremath{{r}}}

\newcommand{\pmdp}{\mathcal{P}}

\newcommand{\ltp}[1][L]{\mathcal{L}}   

\newcommand{\bddstate}[1]{s_{\mathbb{B}}{}}

\newcommand{\tool}[1]{\texttt{#1}\xspace}

\newcommand{\prism}{\tool{PRISM}}
\newcommand{\storm}{\tool{Storm}}

\DeclareRobustCommand{\cpp}
{\valign{\vfil\hbox{##}\vfil\cr
   \textsf{C\kern-.1em}\cr
   $\hbox{\fontsize{\sf@size}{0}\textbf{+\kern-0.05em+}}$\cr}%
}

\selectcolormodel{cmy}

\definecolor{lgrey}{rgb}{0.8,0.8,0.8}
\definecolor{grey}{rgb}{0.5,0.5,0.5}

\definecolor{lightblue}{rgb}{0.8,0.8,1.0}
\definecolor{lightred}{rgb}{1.0,0.8,0.8}
\definecolor{lightgreen}{rgb}{0.8,1.0,0.8}
\definecolor{angrygreen}{cmyk}{0.279,0,0.91,0.08}
\definecolor{lightred}{rgb}{1.0,0.8,0.8}
\definecolor{pink}{rgb}{1.0,0.1,1.0}

\definecolor{prismgreen}{HTML}{009900}
\definecolor{prismred}{HTML}{cc0000}
\definecolor{prismblue}{HTML}{0000cc}

%% file: abstract.tex
In this paper, we propose a method to develop trustworthy reinforcement learning systems. To ensure safety especially during exploration, we automatically synthesize a correct-by-construction runtime enforcer, called a shield, that blocks all actions that are unsafe with respect to a temporal logic specification from the agent. Our main contribution is a new synthesis algorithm for computing the shield online. Existing offline shielding approaches compute exhaustively the safety of all states-action combinations ahead-of-time, resulting in huge offline computation times, large memory consumption, and significant delays at run-time due to the look-ups in a huge database. The intuition behind online shielding is to compute during run-time the set of all states that could be reached in the near future. For each of these states, the safety of all available actions is analysed and used for shielding as soon as one of the considered states is reached. Our proposed method is general and can be applied to a wide range of planning problems with stochastic behavior. For our evaluation, we selected a 2-player version of the classical computer game \snake. The game requires fast decisions and the multiplayer setting induces a large state space, computationally expensive to analyze exhaustively. The safety objective of collision avoidance is easily transferable to a variety of planning tasks.

%% file: introduction.tex
Reinforcement Learning (RL) proved successful in solving complex tasks that are difficult to solve using classic controller design, including applications in computer games~\cite{silver2016mastering}, multi-agent planning~\cite{DBLP:journals/corr/abs-1910-12639}, and robotics~\cite{wang2019learning}. 
RL is able to learn high performance controllers by optimizing objectives expressed via rewards in unknown, stochastic environments. 
Although learning-enabled controllers (LECs) have the potential to outperform classical controllers, safety concerns prevent LECs from being widely used in real-world tasks~\cite{amodei2016concrete}.
In RL, optimal strategies are obtained without prior knowledge about the environment. Therefore, the safety of actions is not known before they are executed. Even after training, there is no guarantee that no unsafe actions are part of the final policy. Having no safety guarantees is unacceptable for safety-critical  areas, such as autonomous driving.

\emph{Shielding}~\cite{DBLP:conf/tacas/BloemKKW15}
 is a runtime enforcement technique to ensure safe decision making.
By augmenting an RL-agent with a shield, at every time step, unsafe actions 
are blocked by the shield and the learning agent is only able to pick a safe action to be sent to the environment. 
Shields are automatically constructed via correct-by-construction formal synthesis methods 
from a model of the environment dynamics and a safety specification.
Consequently, an agent augmented with a shield is guaranteed to satisfy the safety objective as long as the shield is used. 
We model the environment via Markov decision processes (MDPs) which constitute a popular modeling formalism for decision-making under uncertainty~\cite{white1985real,thrun2005probabilistic}.
We assess safety by means of probabilistic \emph{temporal logic constraints}~\cite{BK08} that limit, for example, the probability to reach a set of critical states in the MDP.
For each state and for each action, exact probabilities are computed on how likely it is that executing this action results in a safety violation from the current state.
The shield then blocks all actions whose probability of leading to safety violations exceeds a threshold with respect to an optimally safe action. 

\emph{The problem with offline shielding.}
The computation of an offline shield requires an exhaustive, ahead-of-time safety analysis for all possible state-action combinations. Therefore, the complexity of offline shield synthesis grows exponentially in the state and action dimension, which limits the application of offline shielding to small environments.
Previous work that applied shields in complex, high-dimensional environments
relied on over-approximations of the reachable states and domain-oriented abstractions~\cite{shield_rl,DBLP:conf/cav/AvniBCHKP19}. However, this may result in imprecise safety computations of the shield. This way, the shield may become over-restrictive, hindering the learning agent in properly exploring the environment and finding its optimal 
policy~\cite{DBLP:conf/concur/0001KJSB20}.

\emph{Our solution -- online shielding.} 
Our approach is based on the idea of computing the safety of actions on-the-fly during run time.
In many applications, the learning agent does not have to take a decision at every time step.
Instead, the learning agent only has to make a decision when reaching a \emph{decision state}.
As an example consider a service robot that is controlled to traverse a corridor. 
The agent has time until the service robot reaches the end of the corridor, i.e., the next decision state, to decide on where the service robot should go next. 
The idea of online shielding is to use the time interval between two decision states to
compute the safety of all possible actions for the next decision state. When reaching the next
decision state, this information is used to block unsafe actions of the agent.
While the online safety analysis incurs a runtime overhead, each single computation of the safety of
an action is efficient and parallelizable. 
Thus, in many settings, expensive offline pre-computations and huge shielding databases with costly look-ups are not necessary. 
Since the safety analysis is performed only for decision states that are actually reached,
online shielding is applicable to large, changing, or unknown environments.

In this paper, we 
solve the problem of shielding a controllable RL-agent in an environment shared with other autonomous agents that perform tasks concurrently. 
Some combinations of agent positions are safety-critical, as they e.g. correspond to collisions.
A safety property may describe the probability of reaching such positions (or other safety properties expressible in temporal logic). The task of the shield is to block actions with 
a too high risk of leading to such a state.
In online shielding, the computation of the safety for any action in the next decision state starts as soon as the controllable agent leaves the current decision state.
The tricky part of online shielding in the multi-agent setting is that during the time the RL agent has between two consecutive decisions,
the other agents also change their positions. Therefore, online shielding needs to compute the safety
of actions with respect to all possible movements of the other agents.

Technically, we use MDPs to formalize the dynamics of the agents operating within the environment.
At runtime, we create a small MDP for each decision. 
These MDPs model the immediate future from the viewpoint of the RL. Via model checking, we determine for the next actions the minimal probability of violating safety.
An action is blocked by the shield, if the action violates safety with a probability higher than a threshold relative to the minimal probability. 

\emph{Contributions.} The contributions of this paper comprise (1) the formalization of online shielding, (2) its implementation via probabilistic model-checking including a demonstrator that is available online\footnote{\url{http://www.onlineshielding.at}, accessed: \DTMdisplaydate{2020}{11}{27}{-1}}, and (3) an evaluation of the applicability of online shielding. The implementation and the evaluation apply shielding to a two-player version of the classic computer game \snake. The evaluation demonstrates that shields can be efficiently computed at runtime, guarantee safety, and have the potential to positively influence the learning performance.

\emph{Outline.}
The rest of the paper is structured as follows. \cref{sec:related_work} discusses related work. We discuss the relevant foundations
in Section~\ref{sec:preliminaries}. 
In Section~\ref{sec:problem}, we present the setting and formulate the problem that we address. We present online shielding in Sect.~\ref{sec:shields}, by defining semantics for autonomous agents in the considered setting and defining online shield computations based on these semantics. In Section~\ref{sec:experiments}, we report on the evaluation of online shielding for the classic computer game \snake. \cref{sec:conclusion} concludes the paper with a summary and an outlook on future work.

\subsection{Related Work}
\label{sec:related_work}

Runtime enforcement (RE)~\cite{FalconeP19,RenardFRJM19,DBLP:conf/rv/ZhouGKKL20} covers a wide range of techniques to enforce the correctness of a controller at run-time.
The concept of a correct-by-construction \emph{safety-shield} to enforce such correctness with respect to a temporal logic specification was first proposed in~\cite{DBLP:conf/tacas/BloemKKW15}.
Shields are usually constructed offline by computing a maximally permissive policy that contains all actions that will not violate the safety specification.
Several extensions exist~\cite{DBLP:conf/amcc/BharadwajBDKT19,wu2019shield,DBLP:conf/cav/AvniBCHKP19,DBLP:journals/corr/abs-2010-03842}.
The shielding approach has been shown to be successful in combination with RL~\cite{shield_rl,DBLP:conf/isola/KonighoferL0B20}. 

Jansen et. al.~\cite{DBLP:conf/concur/0001KJSB20} introduced offline shielding with respect to probabilistic safety. Our work on online shielding directly extends their notion of shielding to the online setting.
The offline approach was limited as every action for every state has to be analyzed ahead of time, making the offline approach infeasible for complex environments. Our proposed extension to perform the safety analysis 
online allows the application of shielding in large, high-dimensional environments.

Pranger et al.~\cite{DBLP:journals/corr/abs-2010-03842} proposed \emph{adaptive  shields} to enforce quantitative objectives at run-time.
While the computation of their shields is performed offline,
the authors deal with the consequences of an incorrect or incomplete model that is used for the computation of the shield. During runtime, the authors use abstraction refinement and online probability estimation to update the model and synthesize new shields from the updated models periodically. 

Li et. al.~\cite{DBLP:conf/icra/LiB20} proposed model predictive shielding (MPS).
Given an optimal policy and a safe policy, MPS checks online for each visited state,
whether safety will be maintained using the optimal policy. If not, MPS switches to the safe policy.
In online shielding, we compute the safety of actions before a decision state is visited,
thereby preventing delays at runtime. Furthermore, in online shielding we do not switch
between policies, but evaluate all possible decision to be maximally permissive to the shielded agent.

Safe RL~\cite{garcia2015comprehensive,pecka2014safe,FultonP19} is concerned with providing safety guarantees for learned agents. Our work focusses on the safe exploration~\cite{moldovan2012safe}, we refer to \cite{garcia2015comprehensive} for other types of safe RL. Using their taxonomy, shielding is an instance of ``teacher provides advice''~\cite{DBLP:conf/icml/ClouseU92}, where a teacher with additional information about the system guides the RL agent to pick the right actions. Apprenticeship learning~\cite{DBLP:conf/icml/AbbeelN05} is a closely related variant where the teacher gives (positive) examples and has been used in the context of verification~\cite{DBLP:conf/cav/ZhouL18}. \textsc{Uppaal Stratego} synthesizes safe, permissive policies that are optimized via learning to create controllers for real-time systems~\cite{DBLP:conf/tacas/DavidJLMT15}. 
Some of the work does not assume a model for the environment, making the problem intrinsically harder---and often limiting the safety during exploration.
 We refer to \cite{cheng2019end,fulton2018safe,hasanbeig2018logically,DBLP:conf/tacas/HahnPSSTW19,DBLP:conf/cdc/HasanbeigKAKPL19} for some interesting approaches.

%% file: preliminaries.tex
In this section, we introduce models and properties considered in this paper.

\emph{Sequence and Tuple Notation.}
We denote sequences of elements by $t = e_0 \cdots e_n$ with $\epsilon$ denoting the empty sequence. The length of $t$ is denoted $|t| = n+1$. We use  $t[i] = e_i$ for $0$-based indexed access on tuples and sequences. The notation $t[i \gets e_i']$ represents overwriting of the $i$\textsuperscript{th} element of $t$ by $e_i'$, that is, $t[j] = t[i \gets e_i'][j]$ for all $j\neq i$ and $t[i \gets e_i'][i] = e_i'$. 

A \emph{probability distribution} over a countable set $\distDom$ is a function $\distFunc\colon\distDom\rightarrow[0,1]$ with $\sum_{\distDomElem\in\distDom}\distFunc(\distDomElem)=1$.
$\Distr(\distDom)$ denotes all distributions on $\distDom$. The support of $\distFunc\in\Distr(\distDom)$ is $\supp(\distFunc)=\{x\in\distDom \mid \distFunc(x){>}0\}$.

A \emph{Markov decision process} (MDP) $\MdpInitR$ is a tuple
with a finite set $\states$ of states, a unique initial state $s_0 \in \states$, a finite set $\Act=\{a_1\dots, a_n\}$ of actions, and
a \emph{probabilistic transition function} $\pmdp: \states \times \Act \rightarrow Distr (\states)$.
For all $s \in \states$ the available actions are $\Act(s) = \{a \in \Act | \pmdp(s, a) \neq \bot\}$ and we assume $|\Act(s)| \geq 1$.
A \emph{path} in an MDP $\mdp$ is a finite (or infinite) sequence $\rho=s_0a_0s_1a_1\ldots$ with $\pmdp(s_i, a_i, s_{i+1}) > 0$ for all $i\geq 0$ 
unless otherwise noted.

Non-deterministic choices in an MDP are resolved by a
so-called \emph{policy}. 
For the properties considered in this paper, memoryless deterministic policies are sufficient~\cite{BK08}. These are functions $\pi : \states \rightarrow \Act$ with $\pi(s) \in \Act(s)$.
We denote the set of all memoryless deterministic policies of an MDP by $\Pi$.
Applying a policy $\pi$ to an MDP yields an induced \emph{Markov chain}~ \DtmcInit ~with $\pmdp : \mathcal{S} \rightarrow\Distr(\states)$ where all nondeterminism is resolved.
A \emph{reward function} $\rewFunction \colon \mathcal{S} \times \Act \rightarrow \R_{\geq 0}$ for an MDP adds a reward  to every state $s$ and action $a$ enabled in $s$.

In formal methods, safety properties are often specified as \emph{linear temporal logic} (LTL) properties~\cite{pnueli1977temporal}.
For an MDP $\mdp$, probabilistic model checking~\cite{Kat16,DBLP:conf/lics/Kwiatkowska03}
 employs value iteration or linear programming to compute the probabilities of \emph{all states and actions of the MDP}
 to satisfy an safety property $\varphi$.

Specifically, we compute $\mcresmax{\varphi}{\mdp} \colon \mathcal{S} \rightarrow [0,1]$ or $\mcresmin{\varphi}{\mdp} \colon \mathcal{S} \rightarrow [0,1]$,
which yields for all states the maximal (or minimal) probability over all possible policies to satisfy $\varphi$.
For instance,  for $\varphi$ encoding to reach a set of states $T$,  $\mcresmax{\varphi}{\mdp}(s)$ is the maximal probability to ``eventually'' reach a state in $T$
from state $s\in \mathcal{S}$.

%% file: problem.tex
\paragraph{Setting.}
We consider a setting similar to~\cite{DBLP:conf/concur/0001KJSB20}, 
where one controllable agent, called the \emph{avatar}, and multiple uncontrollable
agents, called \emph{adversaries} operate within an \emph{arena}.
The arena is a compact, high-level
description of the underlying model and captures the dynamic of the agents. 
Any information on rewards is neglected within the arena, since it is not needed for safety computations. 

From this arena, potential agent locations may be inferred. 
Within the arena, the agents perform \emph{tasks} that are sequences of \emph{activities} performed consecutively.

\begin{example}[Robot logistics in a smart factory]
Take a factory floor plan with several corridors with machines.
The adversaries are (possibly autonomous) transporters moving parts within the factory.
The avatar models a specific service unit moving around and inspecting machines where an issue has been raised, while accounting for the behavior of the adversaries.
 Corridors might be too narrow for multiple robots, which poses a safety-critical situation.
A task of the avatar may correspond to inspecting machines. This task consists of several activities that define the path that needs to be taken do visit the machine. 
\end{example}

Formally, an \emph{arena} is a pair $G = (V, E)$, where $V$ is a set of nodes and $E$ is a finite set of  $E \subseteq V\times V$.
An agent’s \emph{location} is defined via the current node $v \in V$.
An edge $(v,v')\in E$ represents an \emph{activity} of an agent. By executing an activity, the agent moves to its next location $v'$.
A \emph{task} is defined as a non-empty sequence $(v_1,v_2) \cdot (v_2,v_3) \cdot (v_3, v_4) \cdots (v_{n-1},v_n) \in E^*$ of connected edges. 
To ease representation, we denote tasks also as sequences of locations $v_1 \cdot v_2 \cdots v_n$. 

The set of tasks available in a location $v \in V$ is given by the function $\Task(v)$.
The set of all tasks of an arena $G$ is denoted by $\Task(G)$.
The avatar is only able to select a next task at a \emph{decision location} in $V_{D}\subseteq V$.
To avoid deadlocks, we require for any decision location $v\in V_{D}$ that $\Task(v) \neq \emptyset$ and for all $v \cdots v' \in Task(v)$ that $v' \in  V_{D}$, i.e., any task ends in another decision locations from which the agent is 
able to decide on a new task.
A safety property may describe that some combinations of agent positions are safety-critical and should not be reached
(or any other safety property from the safety fragment of LTL).

\input{maze}
\begin{example}[Gridworld]
\Cref{fig:maze} shows a simple gridworld with corridors represented by white tiles and walls represented by black tiles. 
A tile is defined via its $(x,y)$ position.
We model this gridworld with an arena $G = (V,E)$ by associating each white tile with a location in $V$ and creating an edge in $E$ for each pair of adjacent white tiles.
Corners and crossings are decision locations, i.e., $V_d = \{(1,1), (1,3), (1,5), (5,1), (5,3), (5,5)\}$. 
At each decision location, tasks define sequences of activities needed to traverse adjoining corridors, e.g., $Task((1,3)) = \{(1,3) \cdot (2,3) \cdot (3,3) \cdot (4,3) \cdot(5,3)$, 
  $(1,3) \cdot (1,2) \cdot (1,1)$,  
  $(1,3) \cdot (1,4) \cdot (1,5)\}$.
\end{example}

\paragraph{Problem.}%
Consider an environment described by an arena as above and a safety specification.
We assume stochastic behaviors for the adversaries, e.g, obtained using RL~\cite{sadigh2018planning,sadigh2016planning}.
In fact, this stochastic behavior determines all actions of the adversaries via probabilities.
The underlying model is then an MDP: the avatar executes an action, and upon this execution the next exact positions (the state of the system) are determined stochastically.

Our aim is to \emph{shield} the decision procedure of the avatar to avoid unsafe behavior regarding the stochastic movements of the adversaries.
\emph{The problem is to compute a shield that prevents the avatar to violate the given safety specification by more than a threshold $\delta$ with respect to the optimal safety probability.}
The safety analysis of actions is performed on-the-fly allowing the avatar to operate within large arenas. 

\begin{example}[Gridworld]
The tile labeled \texttt{A} denotes the location of the avatar and the tile labeled \texttt{E} denotes the position of an adversary. 
Let $(x_A,y_A)$ and $(x_E,y_E)$ be the positions of the avatar and the adversary, respectively.
The given safety property $\varphi = \lnot\mathbf{F}(x_A = x_E \land y_A = y_E)$ states that the agents must not collide. We give more details in Section~\ref{sec:shield_construction} on how to construct a shield for this setting.

\end{example}

%% file: maze.tex
\begin{figure}[t]
\begin{center}
\begin{tikzpicture}[scale=0.95, transform shape]
\draw[step=0.5cm,color=gray] (-1,-1) grid (1.5,1.5);

\node[minimum width = 0.5cm, minimum height = 0.2cm, inner sep = 0pt, outer sep = 0pt] at ( -0.75,+2.0) {1};
\node[minimum width = 0.5cm, minimum height = 0.2cm, inner sep = 0pt, outer sep = 0pt] at ( -0.25,+2.0) {2};
\node[minimum width = 0.5cm, minimum height = 0.2cm, inner sep = 0pt, outer sep = 0pt] at (  0.25,+2.0) {3};
\node[minimum width = 0.5cm, minimum height = 0.2cm, inner sep = 0pt, outer sep = 0pt] at (  0.75,+2.0) {4};
\node[minimum width = 0.5cm, minimum height = 0.2cm, inner sep = 0pt, outer sep = 0pt] at (  1.25,+2.0) {5};

\node[minimum width = 0.2cm, minimum height = 0.5cm] at ( -1.45,1.25) {1};
\node[minimum width = 0.2cm, minimum height = 0.5cm] at ( -1.45,0.75) {2};
\node[minimum width = 0.2cm, minimum height = 0.5cm] at ( -1.45,0.25) {3};
\node[minimum width = 0.2cm, minimum height = 0.5cm] at ( -1.45,-0.25) {4};
\node[minimum width = 0.2cm, minimum height = 0.5cm] at ( -1.45,-0.75) {5};

\node[fill=black, minimum width = 0.2cm, minimum height = 0.2cm] at ( -1.1,+1.6) {};
\node[fill=black, minimum width = 0.5cm, minimum height = 0.2cm] at ( -0.75,+1.6) {};
\node[fill=black, minimum width = 0.5cm, minimum height = 0.2cm] at ( -0.25,+1.6) {};
\node[fill=black, minimum width = 0.5cm, minimum height = 0.2cm] at (  0.25,+1.6) {};
\node[fill=black, minimum width = 0.5cm, minimum height = 0.2cm] at (  0.75,+1.6) {};
\node[fill=black, minimum width = 0.5cm, minimum height = 0.2cm] at (  1.25,+1.6) {};
\node[fill=black, minimum width = 0.2cm, minimum height = 0.2cm] at (  1.6,+1.6) {};

\node[fill=black, minimum width = 0.2cm, minimum height = 0.2cm] at ( -1.1,-1.1) {};
\node[fill=black, minimum width = 0.5cm, minimum height = 0.2cm] at ( -0.75,-1.1) {};
\node[fill=black, minimum width = 0.5cm, minimum height = 0.2cm] at ( -0.25,-1.1) {};
\node[fill=black, minimum width = 0.5cm, minimum height = 0.2cm] at (  0.25,-1.1) {};
\node[fill=black, minimum width = 0.5cm, minimum height = 0.2cm] at (  0.75,-1.1) {};
\node[fill=black, minimum width = 0.5cm, minimum height = 0.2cm] at (  1.25,-1.1) {};
\node[fill=black, minimum width = 0.2cm, minimum height = 0.2cm] at (  1.6,-1.1) {};

\node[fill=black, minimum width = 0.2cm, minimum height = 0.5cm] at ( -1.1,-0.75) {};
\node[fill=black, minimum width = 0.2cm, minimum height = 0.5cm] at ( -1.1,-0.25) {};
\node[fill=black, minimum width = 0.2cm, minimum height = 0.5cm] at ( -1.1, 0.25) {};
\node[fill=black, minimum width = 0.2cm, minimum height = 0.5cm] at ( -1.1, 0.75) {};
\node[fill=black, minimum width = 0.2cm, minimum height = 0.5cm] at ( -1.1, 1.25) {};

\node[fill=black, minimum width = 0.2cm, minimum height = 0.5cm] at (  1.6,-0.75) {};
\node[fill=black, minimum width = 0.2cm, minimum height = 0.5cm] at (  1.6,-0.25) {};
\node[fill=black, minimum width = 0.2cm, minimum height = 0.5cm] at (  1.6, 0.25) {};
\node[fill=black, minimum width = 0.2cm, minimum height = 0.5cm] at (  1.6, 0.75) {};
\node[fill=black, minimum width = 0.2cm, minimum height = 0.5cm] at (  1.6, 1.25) {};

\node[fill=white, minimum width = 0.5cm, minimum height = 0.5cm] at (-0.75,+1.25) {};
\node[fill=white, minimum width = 0.5cm, minimum height = 0.5cm] at (-0.25,+1.25) { };
\node[fill=white, minimum width = 0.5cm, minimum height = 0.5cm] at ( 0.25,+1.25) { };
\node[fill=white, minimum width = 0.5cm, minimum height = 0.5cm] at ( 0.75,+1.25) { };
\node[fill=white, minimum width = 0.5cm, minimum height = 0.5cm] at ( 1.25,+1.25) {A};

\node[fill=white, minimum width = 0.5cm, minimum height = 0.5cm] at (-0.75,+0.75) {};
\node[fill=black, minimum width = 0.5cm, minimum height = 0.5cm] at (-0.25,+0.75) {};
\node[fill=black, minimum width = 0.5cm, minimum height = 0.5cm] at ( 0.25,+0.75) {};
\node[fill=black, minimum width = 0.5cm, minimum height = 0.5cm] at ( 0.75,+0.75) { };
\node[fill=white, minimum width = 0.5cm, minimum height = 0.5cm] at ( 1.25,+0.75) { };

\node[fill=white, minimum width = 0.5cm, minimum height = 0.5cm] at (-0.75,+0.25) { };
\node[fill=white, minimum width = 0.5cm, minimum height = 0.5cm] at (-0.25,+0.25) { };
\node[fill=white, minimum width = 0.5cm, minimum height = 0.5cm] at ( 0.25,+0.25) { };
\node[fill=white, minimum width = 0.5cm, minimum height = 0.5cm] at ( 0.75,+0.25) { };
\node[fill=white, minimum width = 0.5cm, minimum height = 0.5cm] at ( 1.25,+0.25) { };

\node[fill=white, minimum width = 0.5cm, minimum height = 0.5cm] at (-0.75,-0.25) { };
\node[fill=black, minimum width = 0.5cm, minimum height = 0.5cm] at (-0.25,-0.25) { };
\node[fill=black, minimum width = 0.5cm, minimum height = 0.5cm] at ( 0.25,-0.25) { };
\node[fill=black, minimum width = 0.5cm, minimum height = 0.5cm] at ( 0.75,-0.25) { };
\node[fill=white, minimum width = 0.5cm, minimum height = 0.5cm] at ( 1.25,-0.25) { };

\node[circle,minimum size=0.4cm,draw,thick] at (-0.75,-0.75) {};
\node[circle,minimum size=0.3cm,draw,thick] at (-0.75,-0.75) {};

\node[fill=white, minimum width = 0.5cm, minimum height = 0.5cm] at (-0.75,-0.75) {E};

\node[fill=white, minimum width = 0.5cm, minimum height = 0.5cm] at (-0.25,-0.75) { };
\node[fill=white, minimum width = 0.5cm, minimum height = 0.5cm] at ( 0.25,-0.75) { };
\node[fill=white, minimum width = 0.5cm, minimum height = 0.5cm] at ( 0.75,-0.75) { };
\node[fill=white, minimum width = 0.5cm, minimum height = 0.5cm] at ( 1.25,-0.75) { };
\node at (1,+1.25){};
\node at (1,-1.25){};
\end{tikzpicture}

\end{center}
\vspace{-0.5cm}
\caption{Gridworld with avatar A (top right) and an adversary E (bottom left).}
\label{fig:maze}
\end{figure}
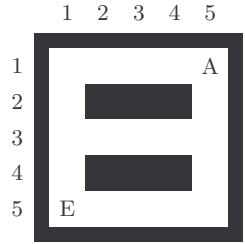

%% file: shields.tex
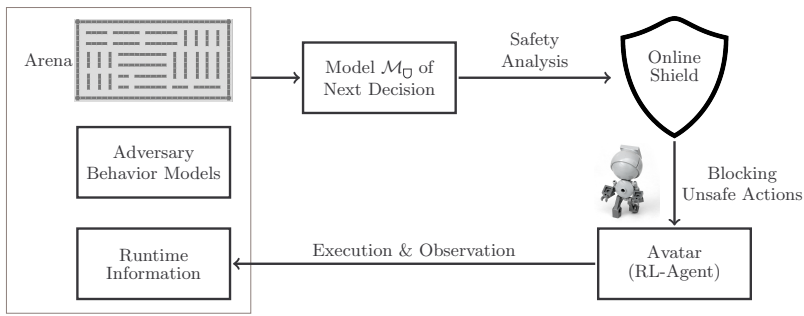
\begin{figure*}[tb]
\centering
\scalebox{0.75}{
\input{pics/tool_sketch}
}
\caption{Workflow of the Shield Construction}
\label{fig:flowchart}
\end{figure*}

In this section, we outline the workflow of online shielding in Figure~\ref{fig:flowchart} and  describe it below. Given an arena and behavior models for adversaries, we define an MDP $\mdp$ that captures all safety-relevant information. 
At runtime, we use current runtime information 
to create sub-MDPs $\online{\mdp}$ of $\mdp$ that model
the immediate future of the agents up to some finite horizon. 
Given such a sub-MDP $\online{\mdp}$ and a safety property $\varphi$, we 
compute via model checking the probability to violate $\varphi$ within the finite horizon
for each task available. The shield then blocks tasks involving a too large risk from the
avatar. 

\subsection{Behavior Models for Adversaries}\label{sec:learning_adv}

The adversaries and the avatar operate within a shared environment, which is represented by an arena $G=(V,E)$, and perform tasks independently.
We assume that we are given a stochastic behavior model of each adversary that determines all task choices
of the respective adversary via probabilities.
The behavior of an adversary is formally defined as follows.

\begin{definition}[Adversary Behavior]
For an arena $G = (V,E)$, we define the behavior $B_i$ of an adversary $i$ as a function $B_i \colon V_D \rightarrow \Distr(\Task(G))$ with $\supp(B_i(v)) \subseteq \Task(v)$.
\end{definition}

Behavior models of adversaries may be derived using domain knowledge or generalized from observations using machine-learning techniques. A potential approach is to observe adversaries in smaller arenas and transfer knowledge gained in this way to larger arenas~\cite{DBLP:conf/concur/0001KJSB20}. Cooperative and truly adverse behavior of adversaries may require to consider additional aspects in the adversary behavior, such as the arena state at a specific point in time. Such considerations are beyond the scope of this paper, since complex adversary behavior generally makes the creation of behavior models more difficult, whereas the online shield computations are hardly affected.

\subsection{Safety-Relevant MDP $\mdp$}

In the following, we describe the safety-relevant MDP $\mdp$ underlying the  agents operating within an arena. This MDP includes non-deterministic choices of the avatar and stochastic behavior of the adversaries.
Note that the safety-relevant MDP $\mdp$ is never explicitly created for online shielding, but is explored on-the-fly for the safety analysis of tasks. 

Let $G=(V,E)$ be an arena, let $\Task$ be a task function for $G$, let $B_i$ with $i\in \{1\ldots m\}$ be the behavior functions of $m$ adversaries, and let the avatar be the zeroth agent. The safety-relevant MDP $\MdpInitR$ models the arena and agents' dynamics as follows. Each agent has a \emph{position} and a \emph{task queue} containing the activities to be performed from the last chosen task.
The agents take turns performing activities from their respective task queue.
If the task queue of an agent is empty, a new task has to be selected.
The avatar chooses non-deterministically, whereas the adversaries choose probabilistically. 

Therefore, $\mathcal{M}$ contains three types of states: (1) states in which the avatar's task queue is empty and the avatar makes a \emph{non-deterministic} decision on its next task, (2) states in which an adversary's task queue is empty and the adversary selects its next task \emph{probabilistically}, and (3) states in which the currently active agent has a non-empty task queue and the agent processes its task queue \emph{deterministically}.

Formally, the \emph{states} $\mathcal{S}=V^{m+1} \times (E^*)^{m+1} \times \{0,\hdots, m\}$ are triples $(v,q,t)$ where $v$ encodes the agent positions, $q$ encodes the task queue states of all agents, and $t$ encodes whose turn it is. There is a unique action $\alpha_\mathrm{adv}$ representing adversary decisions, there is a unique action $\alpha_\mathrm{e}$ representing individual activities (movement along edges), and there are actions for each task available to the avatar, thus $\Act = \{\alpha_\mathrm{adv},\alpha_\mathrm{e}\} \cup \Task(G)$. 

\begin{definition}[Decision State]
Given a  safety-relevant MDP $\mdp$. 
We define the set of \emph{decision states} $\mathcal{S}_D \subseteq \mathcal{S}$ via $\mathcal{S}_D = \{s_D \in \mathcal{S}\mid s_D[1][0] = \epsilon \land s_D[2] = 0\}$,
i.e., it is the turn of the avatar and its task queue is empty.
\end{definition}
This implies that if $s_D\in \mathcal{S}_D$, then $s_D[0][0]$ is a decision location in $V_D$.
A policy for $\mathcal{M}$ needs to define actions only for states in $\mathcal{S}_D$, thereby defining the decisions for the avatar.
At run-time, in each turn each agent performs two steps:
\begin{compactenum}[(1)]
    \item If its task queue is empty, the agent has to select its next task.
    \item The agents performs the next activity of its current task queue.
\end{compactenum}

\paragraph{Selecting a new task.} 
A new task has to be selected in all states $s$ with $s[2] = i$ and $s[1][i] = \epsilon$, i.e, it is the turn of agent $i$ and agent $i$'s task queue is empty.

If $i = 0$, the avatar is in a decision state  $s \in \mathcal{S}_D$. Therefore, $\Act(s) = \Task(s[0][0])$. For each task $t \in \Act(s)$, there is a successor state $s'$ with $s'[1] = s[1][0 \gets t]$, $s'[0] = s[0]$, $s'[2] = s[2]$, and $\pmdp(s,t,s') = 1$. Thus, there is a transition that updates the avatar's task queue 
with the edges of task $t$ with probability one. Other than that, there are no changes. 

If $i\neq 0$, it is an adversary's turn. Therefore, $\Act(s) = \alpha_\mathrm{adv}$. For each $t \in \Task(s[0][i])$, there is a state $s'$ with $s'[1] = s[1][i \gets t]$, $s'[0] = s[0]$, $s'[2] = s[2]$, and $\pmdp(s,\alpha_\mathrm{adv},s') = B_i(s[0][i])(t)$.
There is only a single action, but its outcome is determined stochastically according to the adversary's behavior.

\paragraph{Performing activities.} After potentially selecting a new task, the task queue of agent $i$ is non-empty. We are in a state 
$s'$, where $s'[1][i] = t = (v_i, v_i') \cdot t'$ with $s'[0][i] = v_i$. Agent $i$ moves along the edge $(v_i, v_i')$ deterministically and we increment the turn counter modulo $m+1$, i.e., $\Act(s') = \{\alpha_\mathrm{e}\}$ and $\pmdp(s',\alpha_\mathrm{e},s'') = 1$ with $s''[0] = s'[0][i \gets v_i']$, $s''[1] = s'[1][i \gets t']$, and $s''[2] = s'[2] + 1 \mod m+1$.

\subsection{Sub-MDP $\online{\mdp}$ for Next Decision}
\label{sec:shield_construction}

The idea of online shielding is to compute the safety value of actions in the decision states on 
the fly and block actions that are too risky. 
For infinite horizon properties, the probability to violate safety in the long run is often one and errors stemming from modelling uncertainties may sum up over time~\cite{DBLP:conf/concur/0001KJSB20}.
Therefore, we consider safety relative to a \emph{finite horizon} such that the action values 
(and consequently, a policy for the avatar) carry guarantees for the next steps. 
Explicitly constructing an MDP $\mathcal{M}$ as outlined above yields a very large number of decision states that may be infeasible to check. 
The finite horizon assumption allows us to prune the safety-relevant MDP
and construct small sub-MDPs $\online{\mathcal{M}}$ capturing the immediate future of individual decision states.

In online shielding, we concretely consider runtime situations of being in a state $s_t$, the state visited immediately after the avatar's decision to perform a task $t$. In such situations, we can use the time required to perform $t$ for shield computations for the next decision. 
We create a sub-MDP $\online{\mathcal{M}}$ by determining all states reachable within a finite horizon and use
$\online{\mathcal{M}}$ to check the safety probability of each action (task) available in the next decision
and block unsafe actions. 

\paragraph{Construction of $\online{\mdp}$.}

Online shielding relies on the insight that after deciding on a task $t$, the time required to complete $t$ can be used to compute a shield for the next decision. Thus, we start the construction of the sub-MDP $\online{\mdp}$ for the next decision location $v_D'$ from the state $s_t$ that immediately follows a decision state $s_D$, where the avatar has chosen a task $t \in \Act(s_D)$. The MDP $\online{\mdp}$ is computed with respect to a finite horizon $h$ for $v_D'$. 

By construction, the task is of the form $t = v_D \cdots v_D'$, where $v_D$ is the avatar's current location and $v_D'$ its next decision location. 
While the avatar performs $t$ to reach $v_D'$, the adversaries may perform arbitrary tasks and traverse 
$|t|$ edges, i.e., until $v_D'$ is reached only  adversaries  perform  decisions.
This leads to a set of possible next decision states. We call these states the \emph{first decision states} $S_{FD} \subseteq \mathcal{S_D}$. 
After reaching $v_D'$, both avatar and adversaries decide on arbitrary tasks and all agents traverse $h$ edges. 
This behavior defines the structure of $\online{\mdp}$.

Given a safety-relevant MDP $\MdpInitR$, a decision state $s_D$ and its successor $s_t$ with $s_t[1][0] = t$,
and a finite horizon $h \in \mathbb{N}$ representing a number of turns taken by all agents following the next decision. 
These turns and the (stochastic) agent behavior leading to the next decision are modelled by the sub-MDP $\shielded{\mdp}$.
$\shielded{\mdp}=(\shielded{\mathcal{S}},\shielded{{s_0}},\shielded{\Act},\shielded{\pmdp} )$ is formally constructed as follows. 
The actions are the same as for $\mdp$, i.e., $\shielded{\Act}=\Act$. The initial state is given by $\shielded{{s_0}}=(s_t,0)$.
The states of $\shielded{\mdp}$ are a subset of $\mdp$'s states augmented with the distance from $\shielded{{s_0}}$, i.e., $\online{\mathcal{S}} \subseteq \mathcal{S} \times \mathbb{N}_0$. The distance is measured in terms of the number of turns taken by all agents.

We define transitions and states inductively by: 
\begin{compactenum}[(1)]
\item{\sl Decision Actions.} If $(s,d) \in \online{\mathcal{S}}$, $d < |t| + h$, and there is an $s'\in \mathcal{S}$ such that $\pmdp(s,\act,s') > 0$ and $\act \in \{\act_\mathrm{adv}\} \cup \Task(G)$ then  $(s',d) \in \online{\mathcal{S}}$ and $\shielded{\pmdp}((s,d),\act,(s',d)) = \pmdp(s,\act,s')$. 
\item{\sl Movement Actions.} If $(s,d) \in \online{\mathcal{S}}$, $d < |t| + h$, and there is an $s' \in \mathcal{S}$ such that $\pmdp(s,\act_\mathrm{e},s') > 0$, then  $(s',d') \in \online{\mathcal{S}}$ and $\shielded{\pmdp}((s,d),\act_\mathrm{e},(s',d'))  = \pmdp(s,\act_\mathrm{e},s')$, where $d' = d + 1$ if $s[2] = m$ and $d' = d$ otherwise. 
\end{compactenum} 
Note that a movement of the last agent increases the distance from the initial state. Combined with the fact that every movement action increases the agent index and every decision changes a task queue, we can infer that the structure of $\online{\mdp}$ is a directed acyclic graph. This enables an efficient probabilistic analysis.

By construction, it holds for every state $(s,d) \in \online{\mathcal{S}}$ with $d < |t|$, $s$ is not a decision state of $\mdp$.
The set of first decision states $S_{FD}$ consists of all states $s_{FD} = (s, |t|)$ such that $s_{FD} \in \online{\mathcal{S}}$ with $s[1][0] = \epsilon$ and $s[2] = 0$, i.e., all first decision states reachable from the initial state of $\online{\mdp}$. 
We use $\Task(S_{FD}) = \{t \mid s \in S_{FD}, t \in \Act(s)\}$ to denote the tasks available in these states. $\shielded{\mdp}$ does not define actions and transitions from states $(s,|t| + h) \in \online{\mathcal{S}}$, as their successor states are beyond the considered horizon $h$. 
We have $\Act((s, d)) \neq \emptyset$ for all states at distance $d < |t| + h$ from the initial state.

\subsection{Shield Construction}

A set of unsafe states $T \in \mathcal{S}$ in the safety-relevant MDP should not be reached from any state.
In the finite horizon setting, starting from a decision state $s_D \in \mathcal{S}_D$, states in $T$ should not be reached within the finite horizon $h$.
We evaluate reachability on sub-MDPs $\online{\mdp}$ and use $\online{T} = \{(s,d) \in \online{S} \mid s \in T\}$ to denote the unsafe states that may be reached within the horizon covered by $\online{\mdp}$.
The property $\varphi=\finally \online{T}$ encodes the \textbf{violation} of the safety constraint, i.e., eventually reaching $\online{T}$ within $\online{\mdp}$.
The shield needs to limit the probability to \emph{satisfy} $\varphi$.

Given a sub-MDP $\shielded{\mdp}$ and a set of first decision states $S_{FD}$. For each task $t \in Task(S_{FD})$,
we evaluate $t$ with respect to the minimal probability to satisfy $\varphi$ from the initial state $\shielded{{s_0}}$ when executing $t$ 
by computing $\mcresmin{\varphi}{\online{\mdp}}(\shielded{{s_0}})$. This is formalized with the notion of task-valuations below.

\begin{definition}[Task-valuation]\label{def:actionvalues}
	A \emph{task-valuation} for a task $t$ in a sub-MDP $\online{\mdp}$ with initial state $\shielded{{s_0}}$ and first decision states $S_{FD}$ is given by
	\begin{align*}
 \val{\online{\mdp}}\colon \Task(S_{FD}) \rightarrow [0,1] \text{, } &\text{with } 
 \val{\online{\mdp}}(t) = \mcresmin{\varphi}{\online{\mdp}}(\shielded{{s_0}})\text{,} \\
 &\text{and } \Act(s_{FD}) = \{t\} \text{ for each } s_{FD} \in S_{FD}\text{.}
\end{align*}
The \emph{optimal task-value} for $\online{\mdp}$ is $\optval{\online{\mdp}} =  \min_{t' \in \Task(S_\mathrm{FD})} \val{\online{\mdp}}(t')$.
\end{definition}

A task-valuation is the minimal probability to reach an unsafe state in $T$ from each immediately reachable decision state $s_{FD} \in S_\mathrm{FD}$ weighted by the probability to reach $s_{FD}$. 
When the avatar chooses an optimal task $t$ (with $\val{\online{\mdp}}(t) = \optval{\online{\mdp}}$) as next task in a state $s_{FD}$, $\optval{\online{\mdp}}$ can be achieved if all subsequent decisions are optimal as well.

We now define a shield for the decision states $S_\mathrm{FD}$ in a sub-MDP $\online{\mdp}$ using the task-valuations.
Specifically, a \emph{shield} for a threshold $\delta\in[0,1]$ determines a set of  tasks available in $S_\mathrm{FD}$  that are $\delta$-optimal for the specification $\varphi$. All other tasks are ``shielded'' or ``blocked''.

\begin{definition}[Shield]\label{def:shield}
For task-valuation $\val{\online{\mdp}}$ and a threshold $\delta \in [0,1]$, a \emph{shield for $S_\mathrm{FD}$ in \online{\mdp}} is given by
\begin{align*}
	\shield{\delta}{\online{\mdp}} &\in 2^{\Task(S_\mathrm{FD})} \text{ with } \\
	\shield{\delta}{\online{\mdp}} &= \{ t \in \Task(S_\mathrm{FD}) \mid  \delta \cdot \val{\online{\mdp}}(t) \leq \optval{\online{\mdp}} \}\text{.}
\end{align*}
\end{definition}

Intuitively, $\delta$ enforces a constraint on tasks that are acceptable w.r.t. the optimal probability.
The shield is \emph{adaptive} with respect to $\delta$, as a high value for $\delta$ yields a stricter shield, a smaller value a more permissive shield.
In particularly critical situations, the shield can enforce the decision maker to resort to (only) the optimal actions w.r.t. the safety objective. 
This can be achieved by temporarily setting $\delta = 1$. Online shielding creates shields on-the-fly by constructing sub-MDPs $\online{\mdp}$ and computing task-valuations for all available tasks.

Through online shielding, we transform the safety-relevant MDP $\mdp$ into a \emph{shielded MDP} with which the avatar interacts (which is never explicitly created) that is obtained from the composition of all sub-MDP$\shielded{\mdp}$.
Due to the assumption on the task functions that requires a non-empty set of available tasks in all decision locations
and due to the fact that every decision for shielding is defined w.r.t. an optimal task, the shielded MDP
is deadlock-free. 
In contrast, a shield constructed to enforce safety based on a fixed threshold $\lambda\in[0,1]$ such that only tasks $t$ with $\val{\online{\mdp}}(t)\leq\lambda$ are allowed, may induce deadlocks.
Hence, our notion of online shielding guarantees deadlock-freedom and optimality w.r.t. safety.

\subsection{Optimization -- Updating Shields after Adversary Decisions}
\label{sec:shield_opt}

After the avatar decides on a task, we use the time to complete the task to compute shields based on task-valuations (see \cref{def:actionvalues} and \cref{def:shield}). Such shield computations are inherently affected by uncertainties stemming from stochastic adversary behavior. These uncertainties consequently decrease whenever we observe a concrete decision from an adversary that we considered stochastic in the initial shield computation. 

An optimization of the online shielding approach is to compute a new shield after any decision of an adversary,
if there is enough remaining time until the next decision location.
Suppose that after visiting a decision state, we computed a shield based on $\shielded{\mdp}$.
While moving to the next decision state, an adversary decides on a new task and we observe the concrete state $s$.
We can now construct a new sub-MDP $\online{\mdp'}$ using $\shielded{{s_0'}}=s$ as initial state, thereby resolving a stochastic decision between the original initial state $\shielded{{s_0}}$ and $\shielded{{s_0'}}$. 
Using $\online{\mdp'}$, we compute a new shield for the next decision location.

The facts that the probabilistic transition function of $\online{\mdp}$ does not change during updates and that we consider safety properties enable a very efficient implementation of updates. For instance, if value iteration is used to compute task-valuations, we can simply change the initial state and reuse computations from the initial shield computation. 
Note that if a task is completely safe, i.e., tasks with a valuation of zero, the value of this task will not change under a re-computation, since the task is safe under any sequence of adversary decisions. 

%% file: pics/tool_sketch.tex
\tikzstyle{doc}=[
draw,
very thick,
align=center,
color=black,
text width=2.6cm,
minimum height=1.25cm,
inner sep=1.5pt,
outer sep=1.5pt,
text centered,
on grid,
fill=white
]

\begin{tikzpicture}

	\node[text centered, doc] (adversary_behavior) {Adversary Behavior Models};
	
	\node[inner sep=0pt, rectangle, above=0.4cm of adversary_behavior] (arena) {
 	{
 	\includegraphics[width=2.6cm+6pt]{
 	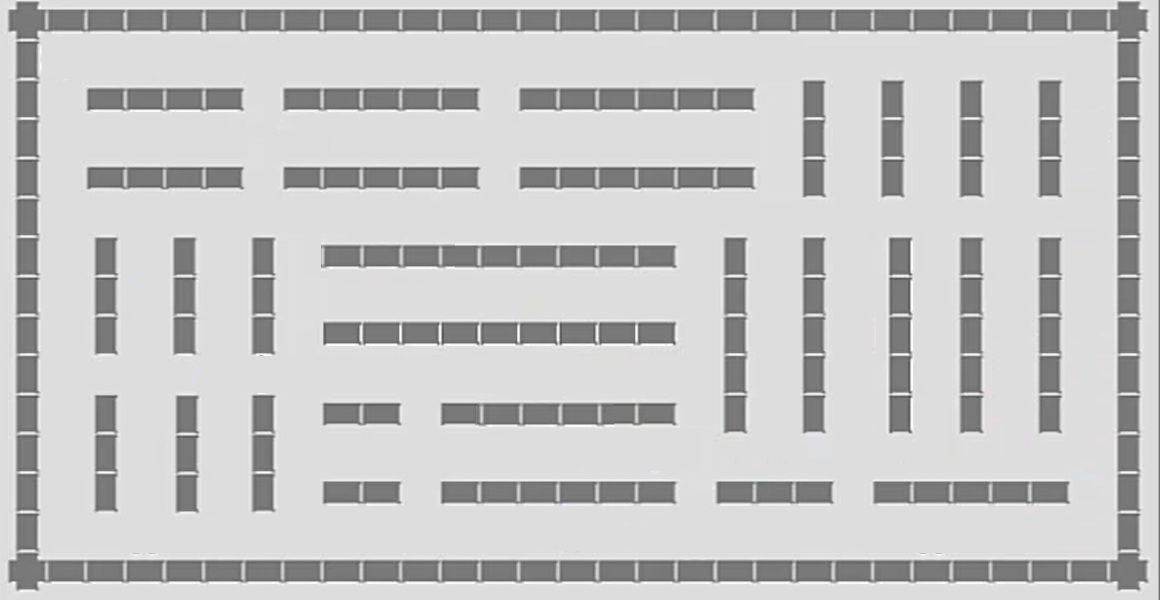}
 	}
	};
	\node[left=-0.20cm of arena] (arenalabel) {Arena};
	
	\node[inner sep=0pt, rectangle, doc, below=1.8cm of adversary_behavior] (runtime_information) {\shortstack{Runtime\\Information}};

	\node[inner sep=0pt, rectangle, doc, above right=1.5cm and 4cm of adversary_behavior] (model_next_decision) {Model $\online{\mdp}$ of Next Decision};
\node[inner sep=1pt, right=2.7cm of model_next_decision] (shield_pic) {{{\includegraphics[scale=0.12]{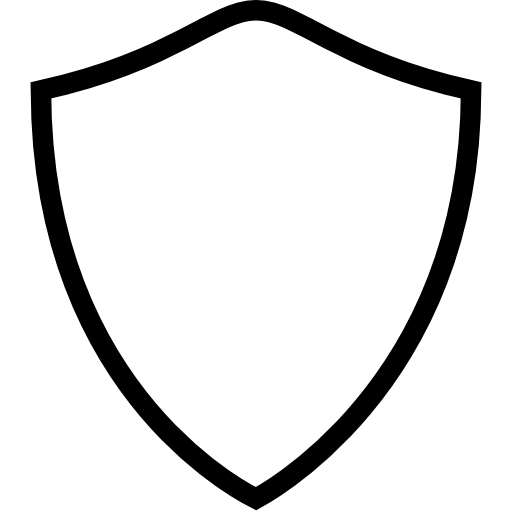}}}};
\node[inner sep=1pt, below=0.6cm of shield_pic.north] {\shortstack{Online\\Shield}};

\node[doc] at (runtime_information-|shield_pic)(shield_rl) {\shortstack{Avatar \\ (RL-Agent)}};

	\node[fit=(adversary_behavior)(arenalabel) (arena) (runtime_information), draw=gray, inner sep=0.2cm] (info_for_next_dec) {};

 	\draw[->] (info_for_next_dec.east|-model_next_decision)[very thick] -- node[auto] {} (model_next_decision);
 	\draw[->] (model_next_decision)[very thick] -- node[auto] {\shortstack{Safety \\ Analysis}} (shield_pic);
 	
	\draw[->] (shield_pic)[very thick] -- node[auto, right](edge_to_shield_rl) {\shortstack{Blocking \\ Unsafe Actions}} (shield_rl);
	
	\draw[->] (shield_rl)[very thick] -- node[auto, above] {\shortstack{Execution \& Observation}} (runtime_information);
	
	\node[inner sep=0pt, rectangle, left=0cm of edge_to_shield_rl] (avatar) {{\includegraphics[bb= 15 10 800 780,scale=0.05]{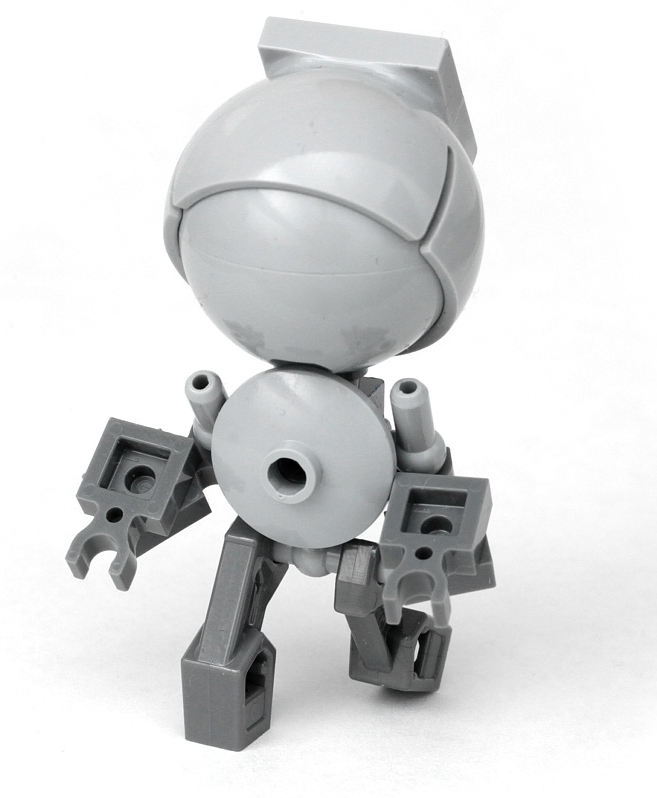}}};

\end{tikzpicture}

%% file: experiments.tex
\paragraph{2-Player \snake.}
\begin{figure}[t]
\centering
\includegraphics[width=0.45\textwidth]{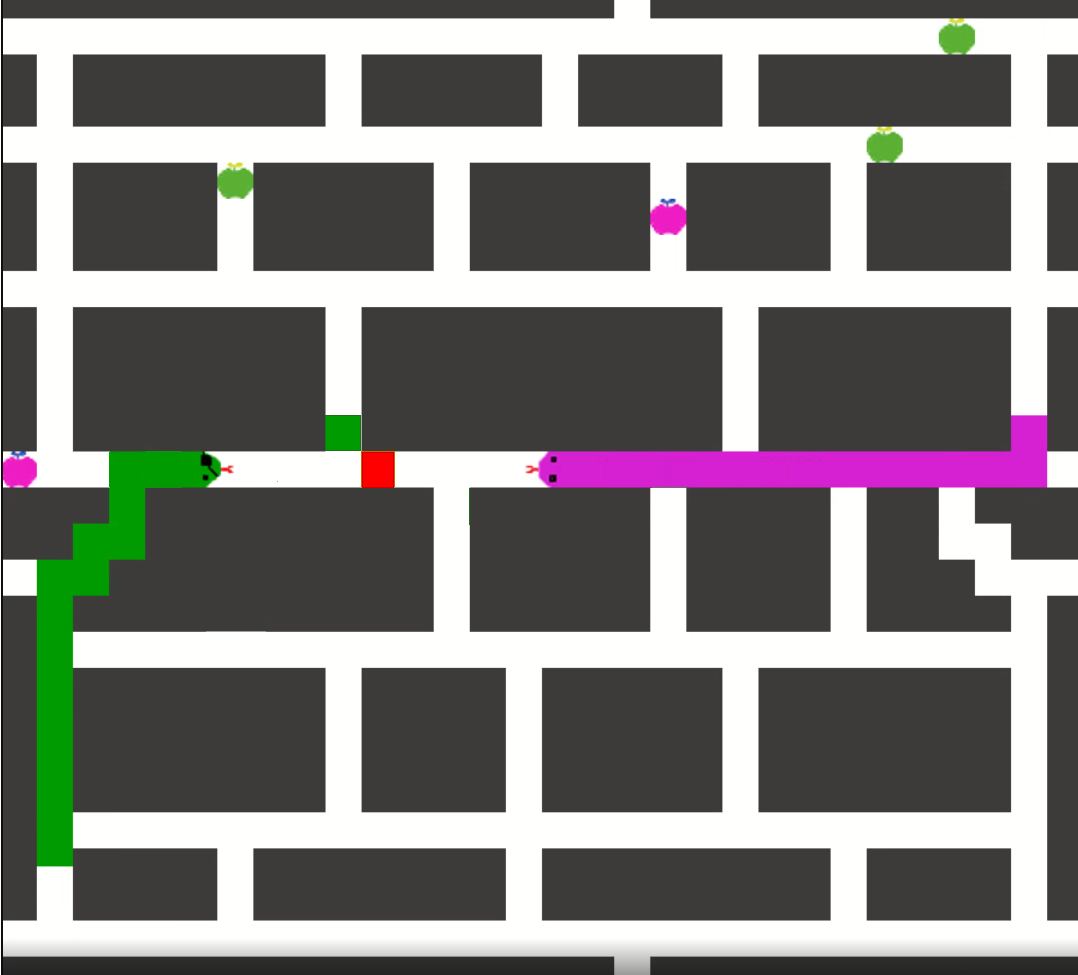}
\caption{A screenshot of the \snake game with color-coded shield display.}
\label{fig:sceenshot1}
\end{figure}
We implemented online shielding for a two-player version of the classic computer game \snake. 
We picked the game because it requires fast decision making during runtime, and
provides in an intuitive and fun setting to show the potential of shielding such that it can potentially be used for teaching formal methods. 
Figure~\ref{fig:sceenshot1} shows a screenshot of the 2-Player \snake game on the map that was used for the experiments. In the game, each player controls a snake of a different color. Here, the green snake is controlled by the avatar (the RL-agent) and the purple snake by the adversary. 
The goal for each player is to eat five randomly positioned apples of their own color.
The score for the green snake (the avatar) is positively
affected (+10) by collecting a green apple and by wins of the avatar (+50), i.e., if it collects all green apples before the adversary snake collects all purple apples.
If a snake has a collision, the snake loses. In case that the heads of both snakes collide, the avatar loses.

We implemented a shield to protect the avatar snake from collisions with the adversary snake. The shield computes online the minimal probability that taking the next corridor will lead to a collision.
In the game, we indicate the risk of taking a corridor from low to high by the colors green, yellow, orange, red, as shown in Figure~\ref{fig:sceenshot1}. 

\begin{figure}[t]
\centering
\includegraphics[width=0.59\textwidth]{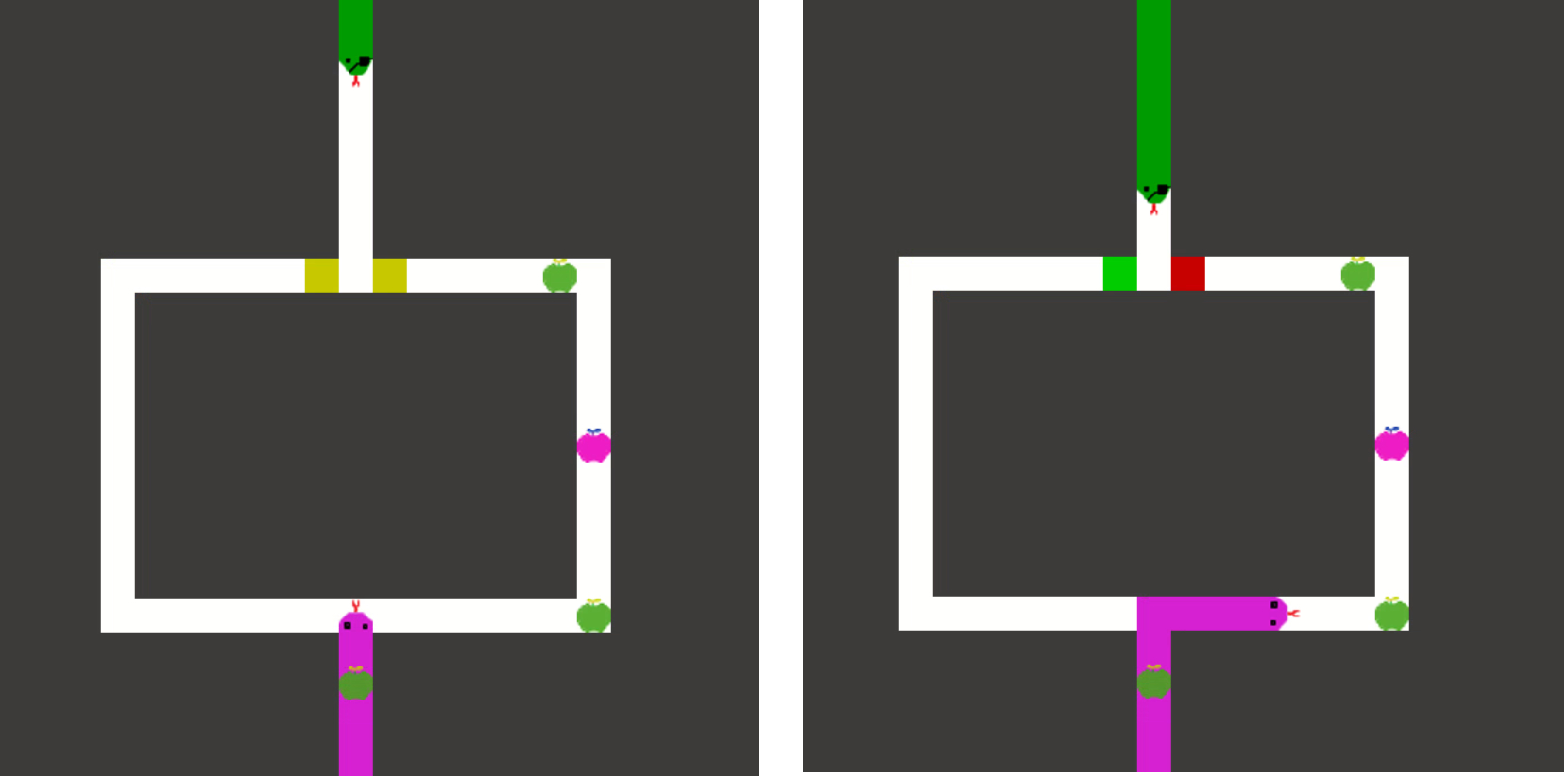}
\caption{Screenshots from the \snake game to demonstrate Recalculation.}
\label{fig:sceenshot_shield_and_recalc}
\end{figure}
We also implemented the optimization to recalculate the shield after a decision made by the adversary snake.
Figure~\ref{fig:sceenshot_shield_and_recalc} contains two screenshots of the game on a simple map to demonstrate 
the effect of a shield update. In the left figure, the available tasks of the green snake 
are picking the corridor to the left or the corridor to the right.
Both choices induce a risk of a collision with the purple snake. After the decision of 
the purple snake to take the corridor to its right hand-side, the shield is updated and the 
safety values of the corridors change.

\paragraph{Experimental Set-up.}
The Python-based implementation can be found at \url{http://onlineshielding.at} along with videos, evaluation data and a Docker image that enables easy experimentation.
For shield computations, we use the probabilistic model checker \storm~\cite{DJKV17} and its Python interface. We use the \prism~\cite{DBLP:conf/cav/KwiatkowskaNP11} language to represent MDPs and domain-specific optimizations to efficiently encode agents and tasks, that is, snakes and their movements.
Reinforcement learning is implemented via approximate Q-learning~\cite{sutton1998reinforcement}
with the feature vector denoting the distance to the next apple.
The Q-learning uses the learning rate $\alpha = 0.1$ and the
discount factor $\gamma = 0.5$ for the Q-update and an $\epsilon$-greedy exploration policy with $\epsilon$ = 0.6.
The pygame\footnote{\url{https://www.pygame.org/}, accessed \DTMdisplaydate{2020}{11}{27}{-1}} library is used to implement the game's interface and logic. 
All experiments have been performed on a computer with an Intel\textregistered{}Core\texttrademark{} i7-4700MQ CPU with 2.4 GHz, 8 cores and 16 GB RAM.

\paragraph{Evaluation Criteria.}
We report on two types of experiments: (1) the time required to compute shields relative to the computation horizon and (2) the performance of shielded reinforcement learning compared to unshielded reinforcement learning measured in terms of gained reward. The experiments on computation time indicate how many steps shielding can look ahead within some given time. The experiments on learning performance demonstrate the effect of shielding.

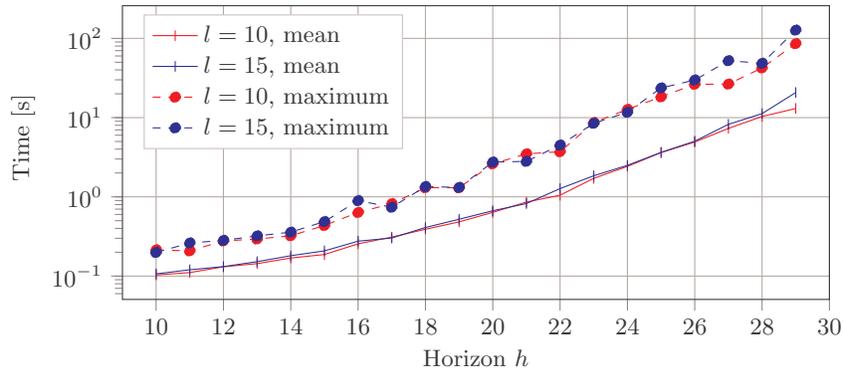
\begin{figure}[t]
    \centering
    \input{pics/snake_time}
    \caption{Shield computation time for varying horizon values and snake lengths.}
    \label{fig:eval_time}
\end{figure}

\paragraph{Computation Time Measurements.}
When playing the game on the map illustrated in Figure~\ref{fig:sceenshot1}, we measured the time to compute shields, i.e., the time to construct sub-MDPs $\shielded{\mdp}$ and to compute the safety values.
We measured the time of $200$ such shield computations and report the the maximum computation times and the mean computation times.
Figure~\ref{fig:eval_time} presents the results for two different snake lengths $l \in \{10, 15\}$ and
different computation horizons $h \in \{10, 11, \dots, 29\}$. 
The x-axis displays the computation horizon $h$ and the y-axis displays the computation time in seconds in logarithmic scale.

We can observe that up to a horizon $h$ of $17$, all computations take less than one second, even in the worst case. Assuming that every task takes at least one second, we can plan ahead by taking into account safety hazards within the next $17$ steps. A computation horizon of $20$ still requires less than one second on average and about $3$ seconds in the worst case. Horizons in this range are often sufficient, as we demonstrate in the next experiment by using $h=15$. 

We compare our timing results with a similar case study presented by Jansen et. al.~\cite{DBLP:conf/concur/0001KJSB20}.
In a similar multi-agent setting on a comparably large map, 
the decisions of the avatar were shielded using an offline shield
with a finite horizon of $10$. The computation time to compute the offline shield
was about \emph{6 hours} on a standard notebook.
Note, that although the setting has four adversaries, the offline computation was performed for one adversary and the results were combined for several adversaries online.

Furthermore, \cref{fig:eval_time} shows that the snake length affects the computation time only slightly. This observation supports our claim that online shielding scales well to large arenas, i.e., scenarios where the safety-relevant MDP $\mdp$ is large. Note that the number of game configurations grows exponentially with the snake length (assuming a sufficiently large map), as the snake's tail may bend in different directions at each crossing. 

The experiments further show that the computation time grows exponentially with the horizon. 
Horizons close to $30$ may be advantageous in especially safety-critical settings, such as factories with industrial robots acting as agents. 
Since individual tasks in a factory may take minutes, online shielding would be feasible, as worst-case computation times are in the range of minutes. However, offline shielding would be clearly infeasible due to the average computation time of more than $10$ seconds that would be required for all decision states.

\paragraph{RL with Online Shielding.}
\label{sec:rewards_shielding_experiments}
\begin{figure}[t]
    \centering
\input{pics/reward_plot}
    \caption{Reward gained throughout learning for shielded and unshielded RL.}
    \label{fig:reward_plots}
\end{figure}
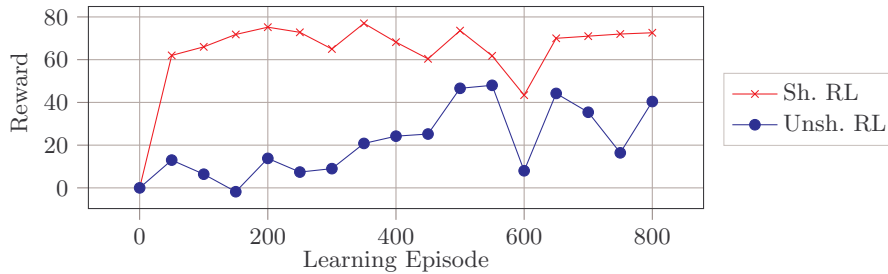

Figure~\ref{fig:reward_plots} shows plots of the reward gained during learning in the shielded and in the unshielded case. The online shield uses a horizon of $h=15$. The y-axis displays the reward and the x-axis displays the learning episodes, where one episode correspond to one play of the \snake game. The reward has been averaged over 50 episodes for each data point. 

The plot demonstrates that shielding improves the gained reward significantly. By blocking unsafe actions, 
the avatar did not encounter a single loss due to a collision. 
For this reason, we see a consistently high reward right from the start of the learning phase. 
In the execution phase, shielded RL manages to win about $96\%$ of all plays, where as unshielded RL wins only about $54\%$.

%% file: pics/snake_time.tex
\begin{tikzpicture}
\begin{axis}[legend cell align={left},
legend entries={{$l=10$, mean},{$l=15$, mean},{$l=10$, maximum},{$l=15$, maximum}},
legend style={at={(0.03,0.97)}, anchor=north west, draw=white!80.0!black},
tick align=outside,
tick pos=left,
x grid style={lightgray!92.02614379084967!black},
xlabel={Horizon $h$},
xlabel style = {below=2mm},
ylabel style = {above=2mm},
title style = {above=3mm},
xmajorgrids,
xmin=9, xmax=30,
y grid style={lightgray!92.02614379084967!black},
ylabel={Time [s]},
ymajorgrids,
width=0.9\textwidth,
height=0.45\textwidth,
x label style={at={(axis description cs:0.5,-0.09)},anchor=north},
ymode=log]

\addplot[color=red, style = solid, mark = |] coordinates { 
(10,0.10339624687447213)
(11,0.11077590154571226)
(12,0.1310319801006699)
(13,0.14368726900283946)
(14,0.16911989896558224)
(15,0.18721914495923556)
(16,0.2554155680484837)
(17,0.3106202617267263)
(18,0.38990108019876063)
(19,0.48311986114567845)
(20,0.6370980332864565)
(21,0.8593174993235152)
(22,1.0488577481624088)
(23,1.713255172832287)
(24,2.4239968288896487)
(25,3.622418735098327)
(26,4.922773461265024)
(27,7.30723524538509)
(28,10.343160685923358)
(29,13.030479521512461)
};
\addplot[color=blue, style = solid, mark = |] coordinates { 
(10,0.10665499687544071)
(11,0.12013743004150455)
(12,0.13175734899123198)
(13,0.1521771593161975)
(14,0.1807794839469716)
(15,0.20805640439968556)
(16,0.27587092314293843)
(17,0.30226686300506117)
(18,0.4104779734439217)
(19,0.5215225122190895)
(20,0.6681664935959271)
(21,0.8234064633597155)
(22,1.271768888492079)
(23,1.8452586279041134)
(24,2.4979558239213655)
(25,3.6438199609945876)
(26,5.036405216602143)
(27,8.23255051202228)
(28,11.235037990765704)
(29,20.738799952500557)
};
\addplot[color=red, style = dashed, mark = *] coordinates { 
(10,0.21436117397388443)
(11,0.2079704569769092)
(12,0.2780317129800096)
(13,0.2936996130156331)
(14,0.3228294490254484)
(15,0.4350784639827907)
(16,0.6355329009820707)
(17,0.8210048309992999)
(18,1.3053794450243004)
(19,1.3036258540232666)
(20,2.610773002030328)
(21,3.4872464640066028)
(22,3.7140747309895232)
(23,8.657626482017804)
(24,12.714245119015686)
(25,18.282771444995888)
(26,26.35125045699533)
(27,26.49520757497521)
(28,42.40304919099435)
(29,85.78463311999803)
};
\addplot[color=blue, style = dashed, mark = *] coordinates { 
(10,0.19867584604071453)
(11,0.2620728629990481)
(12,0.282208206015639)
(13,0.32163305598078296)
(14,0.35842895798850805)
(15,0.48769047000678256)
(16,0.8963309100363404)
(17,0.7393665809649974)
(18,1.349632297991775)
(19,1.3153385600307956)
(20,2.748471908038482)
(21,2.806087204022333)
(22,4.5055409540073015)
(23,8.473888397973496)
(24,11.683173689001705)
(25,23.68352826498449)
(26,29.85266632999992)
(27,52.281756129988935)
(28,48.5033698239713)
(29,127.10364941897569)
};

\end{axis}
\end{tikzpicture}

%% file: pics/reward_plot.tex
\begin{tikzpicture}
\begin{axis}[legend cell align={left},
legend entries={{Sh. RL},{Unsh. RL}},
legend style={at={(1.32,0.31)}, anchor= south east, draw=white!80.0!black},
tick align=outside,
tick pos=left,
x grid style={lightgray!92.02614379084967!black},
xlabel={Learning Episode},
xlabel style = {below=2mm},
ylabel style = {above=2mm},
title style = {above=3mm},
xmajorgrids,
y grid style={lightgray!92.02614379084967!black},
ylabel={Reward},
ymajorgrids,
width=0.8\textwidth,
height=0.35\textwidth,
x label style={at={(axis description cs:0.5,-0.09)},anchor=north},
]

\addplot [color=red, style = solid, mark = x] coordinates { 
(0,0)
(50,62.0)
(100,66.0)
(150,71.8)
(200,75.2)
(250,72.8)
(300,65.0)
(350,77.0)
(400,68.2)
(450,60.4)
(500,73.6)
(550,61.8)
(600,43.4)
(650,70.0)
(700,71.0)
(750,72.0)
(800,72.6)
};

\addplot [color=blue, style = solid, mark = *] coordinates { 
(0,0)
(50,13.0)
(100,6.4)
(150,-1.8)
(200,13.8)
(250,7.4)
(300,9.0)
(350,20.8)
(400,24.2)
(450,25.2)
(500,46.6)
(550,48.0)
(600,8.0)
(650,44.2)
(700,35.4)
(750,16.4)
(800,40.4)
};
\end{axis}
\end{tikzpicture}

%% file: conclusion.tex
Online shielding is an efficient approach to enforce safe behavior of autonomous agents operating within a stochastic environment. The approach exploits the time required to complete tasks to model and analyze the immediate future w.r.t. a safety property. For every decision at runtime, we model the current state of the environment and the behavior of the agents in the form of MDPs. Based on these MDP models, we employ probabilistic model-checking to evaluate every action possible in the next decision. In particular, we determine the probability of unsafe behavior following each possible choice. This information is used by shields to block unsafe actions, i.e., actions leading to a safety violations with probability exceeding a threshold relative to minimal probability of safety violations.

For future work, we plan to investigate the application of online shielding in other settings, such as decision making in robotics and control. Another interesting extension would be to incorporate quantitative performance measures in the form of rewards and costs into the computation of the online shield, as previously demonstrated in an offline manner~\cite{DBLP:conf/cav/AvniBCHKP19} and in a hybrid approach~\cite{DBLP:journals/corr/abs-2010-03842}, where runtime information was used to learn the environment dynamics.